\documentclass[aps,prl,twocolumn,groupedaddress]{revtex4}

\usepackage{graphicx}
\usepackage{dcolumn}
\usepackage{bm}
\usepackage{subfigure}

\begin{document}


\title{Glassy Freezing and Long-Range Order in the $3D$ Random Field $XY$ Model}


\author{Ronald Fisch}
\email[]{ron@princeton.edu}
\affiliation{382 Willowbrook Dr.\\
North Brunswick, NJ 08902}


\date{\today}

\begin{abstract}
Monte Carlo studies of the 3D random field $XY$ model on simple
cubic lattices of size $64^3$, using two different isotropic
random-field probability distributions of moderate strength,
show a glassy freezing behavior above $T_c$, and long-range
order below $T_c$, consistent with our earlier results for weaker
and stronger random fields.  This model should describe random
pinning in vortex lattices in type-II superconducting alloys,
charge-density wave materials, and decagonal quasicrystals.

\end{abstract}

\pacs{75.10.Nr, 74.25.Uv, 64.60.De, 61.44.-n}

\maketitle


The random-field XY model (RFXYM) was first studied forty years ago,\cite{Lar70}
as a toy model for random pinning effects in the vortex lattice of a type-II
superconductor in an external magnetic field.  For fixed-length classical spins
the Hamiltonian of the RFXYM is
\begin{equation}
  H ~=~ - J \sum_{\langle ij \rangle} \cos ( \phi_{i} - \phi_{j} )
  ~-~ \sum_{i} h_i \cos ( \phi_{i} - \theta_{i} )  \, .
\end{equation}
Each $\phi_{i}$ is a dynamical variable which takes on values
between 0 and $2 \pi$. The $\langle ij \rangle$ indicates here a sum
over nearest neighbors on a simple cubic lattice of size $L \times L
\times L$. We choose each $\theta_{i}$ to be an independent
identically distributed quenched random variable, with the flat
probability distribution
\begin{equation}
  P ( \theta_i ) ~=~ 1 / 2 \pi   \,
\end{equation}
for $\theta_i$ between 0 and $2 \pi$.  The probability distribution
$P ( h_i )$ is independent of $\theta_i$.  This Hamiltonian is closely
related to models of incommensurate charge density waves,\cite{GH96,Fis97}
which include decagonal and other axial quasicrystals\cite{SJSTAT98} as
special cases.  In all of these cases, the physical origin of the ``random
field" is the fact that any real sample contains a finite density of point
defects, {\it i.e.} atoms sitting in places where they would not be in the
ideal structure.

We will use the term ``sample" to denote a Hamiltonian with a particular
set of the $h_i$ and $\theta_{i}$ variables.  A thermal average of some
quantity over a single sample will be denoted here by angle brackets,
$\langle ~ \rangle$, and an average over all samples will be denoted by
square brackets, $[ ~ ]$.

Larkin\cite{Lar70} replaced the spin-exchange term of the Hamiltonian
with a harmonic potential, so that each $\phi_{i}$ is no longer restricted
to lie in a compact interval.  This approximation, which neglects mode
coupling, is strictly valid only in the limit $n \to \infty$, where $n$
is the number of components for each spin.  Larkin then argued that, for
any probability distribution $P ( h_i )$ which allows non-zero values for
$h_i$, this model has no long-range ordered phase on a lattice whose
spatial dimension $d$ is less than or equal to four.

A more intuitive derivation of this result was given by Imry and
Ma,\cite{IM75} who assumed that the increase in the energy of an $L^d$
lattice when the order parameter is twisted at a boundary scales as
$L^{d - 2}$.  This was later justified\cite{AIM76,PS82} by showing that
within a perturbative $\epsilon$ expansion one finds the phenomenon of
``dimensional reduction".  Term by term for $T > T_c$, the critical
exponents of any $d$-dimensional $O(n)$ random-field model appear to be
identical to those of an ordinary $O(n)$ model of dimension $d - 2$.

However, reservations about the correctness of applying the $n = \infty$
results for finite $n$, {\it i.e.} the substitution of the harmonic
potential for the exchange term, were soon expressed by
Grinstein.\cite{Grin76}  For the Ising ($n = 1$) case, dimensional
reduction was shown rigorously to be incorrect.\cite{Imb84,BK87}

One problem with dimensional reduction for $n > 1$ is that the
perturbation theory can only be calculated for $T > T_c$, while the
twist energy of the order parameter is only defined for $T < T_c$.
Another, more interesting problem for $n > 1$ is associated with the
question of properly defining the order parameter.  As discussed, for
example, by Griffiths,\cite{Grif66} there is more than one way to do
this.  For the random-field model, there are several different types of
two-point correlation functions.\cite{Grin76}

When $n = 2$, one may use the vector order parameter $\vec{\bf M}$, defined
by
\begin{equation}
  \vec{\bf M} ~=~  N^{-1} \sum_{i} \cos ( \phi_{i} ) \hat{x} ~+~
  \sin ( \phi_{i} ) \hat{y}   \, ,
\end{equation}
where $N = L^d$ is the number of spins.  Alternatively, one can choose the
scalar order parameter $M^2$, defined by
\begin{equation}
  M^2 ~=~  N^{-2} \sum_{i,j} \cos ( \phi_{i} ) \cos ( \phi_{j} ) ~+~
  \sin ( \phi_{i} ) \sin ( \phi_{j} )   \, .
\end{equation}
It is very awkward to calculate $\vec{\bf M}$, since its value will vary wildly
from sample to sample.  Thus we expect that $[\vec{\bf M}]$ will be zero, even
though $\langle \vec{\bf M} \rangle$ for each sample may become large at low
temperature.  On the other hand, we anticipate that for weak disorder the
probability distribution for $M^2$ will become narrow as we take $L \to \infty$.
Thus it is the quantity $\langle M^2 \rangle$, and not the quantity
$\langle \vec{\bf M} \rangle$, which is the natural carrier of long-range order
in the presence of the random field.

It is also necessary to consider the possibility that the thermal average
$\langle M^2 \rangle$ for a typical finite sample may become large as $T$ is
lowered, even though $\langle \vec{\bf M} \rangle$ remains small.  This will
occur when the energy required for reorienting $\langle \vec{\bf M} \rangle$ is
nonzero, but not large compared to $T$.  If this were to occur, then there would
be no local order parameter, and thus no terms in the Landau-Ginsburg free-energy
functional which depended on gradients of $\langle \vec{\bf M} \rangle$.

Monte Carlo calculations for Eqn.~(1) on simple cubic lattices were performed
some time ago by Gingras and Huse\cite{GH96} and Fisch\cite{Fis97}.  More
recently, additional calculations were carried out.\cite{Fis07}  The results
of these studies, which used several different choices for the probability
distribution $P ( h_i )$, indicate that for typical samples the order parameter
$\langle M^2 \rangle$ becomes positive at some positive temperature $T_c$, as
long as the random fields are not too strong.  However, the precise nature of
what happens near $T_c$ remained unclear.  In this work we report additional
studies using new choices of $P ( h_i )$, which display phenomena which were
not seen clearly in the earlier studies.

The computer program which was used was an enhanced version of the one used
before.\cite{Fis07}  The data were obtained from $L \times L \times L$
simple cubic lattices with $L = 64$ using periodic boundary conditions.
Some preliminary studies for smaller values of $L$ were also done.  The
program approximates $O(2)$ with $Z_{12}$, a 12-state clock model.  It was
modified to enable the study of random-field probability distributions of
the form
\begin{equation}
  P ( h_i ) ~=~  (1 - p) \delta ( h_i - h_r ) ~+~ p \delta ( h_i )  \,   .
\end{equation}
This modification makes the program run a few percent slower than before.
In this work we study two cases.  Type A samples have $h_r = 1.5$ with
$p = 0$, and Type B samples have $h_r = 2$ with $p = 0.5$.  For both
of these cases we find $T_c / J$ near 1.5, which was suggested by Gingras
and Huse\cite{GH96} to be optimal for this sort of calculation.

On a simple cubic lattice of size $L \times L \times L$, the magnetic
structure factor, $S (\vec{\bf k})$, for $n = 2$ spins is
\begin{equation}
  S (\vec{\bf k}) ~=~  L^{-3} \sum_{ i,j } \cos ( \vec{\bf k} \cdot
  \vec{\bf r}_{ij}) \langle \cos ( \phi_{i} - \phi_{j}) \rangle  \,   ,
\end{equation}
where $\vec{\bf r}_{ij}$ is the vector on the lattice which starts
at site $i$ and ends at site $j$.  Thus, setting $\vec{\bf k}$ to zero yields
\begin{equation}
  S ( 0 ) / L^3 ~=~ \langle M^2 ( L ) \rangle  \,   .
\end{equation}
As we will see, the extrapolation of $\langle M^2 ( L ) \rangle$ to $L =
\infty$ is not trivial.

In this work, we present results for the average over angles of
$S (\vec{\bf k})$, which we write as $S ( k )$.  Eight different $L = 64$
samples of the random fields $\theta_i$ were used for each $P ( h_i )$.
The same samples of random fields were used for all values of $T$.

The Monte Carlo calculations were performed using both hot start
({\it i.e.} random) initial conditions followed by slow cooling, and also
cold start ({\it i.e.} ferromagnetic) initial conditions followed by
slow warming.  For the Type A samples three different cold start initial
conditions were used, equally spaced around the circle.  For the Type B
samples, where the average random field is somewhat weaker, only two cold
start initial conditions were used.

\begin{figure}
\includegraphics[width=3.4in]{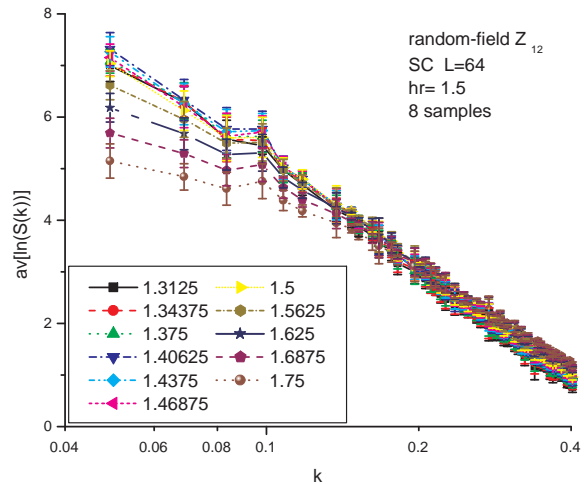}
\caption{\label{Fig.1}(color online) Average over 8 samples of $\ln(S(k))$
for $64 \times 64 \times 64$ lattices with $h_r = 1.5$ and $p = 0$ at
various temperatures. The error bars indicate approximately three standard
deviation statistical errors, and the $x$-axis is scaled logarithmically.}
\end{figure}

In Fig.~1 we show data for the average over our eight Type A samples of
$\ln(S(k))$ for $T / J$ between 1.3125 and 1.75.  The data for $T / J \ge
1.4375$ are taken from the hot start runs, and the data for lower values
of $T / J$ are taken from the cold start run which gave the lowest average
value for the energy, $\langle E \rangle$.  When $T = 1.4375$ or more, all
runs for a given sample converge in a rather short time to give similar
values of $\langle E \rangle$ and $\langle \vec{\bf M} \rangle$.  However,
for lower values of $T / J$ the hot start runs remain stuck in metastable
states.

An attempt was made to drive the hot start runs into equilibrium by deep
undercooling, followed by slow warming.  This thermal cycling technique
had proven successful earlier,\cite{Fis07} for the weak random field case
of $h_r = 1.0$ with $p = 0$.  For the cases studied here, however, a
single undercooling was only partially successful in driving the samples
to equilibrium.  The author believes that repeating cycling would have
worked successfully.  However, since a detailed study of the
nonequilibrium behavior was not the purpose of the work described here,
this was not attempted.

Somewhat surprisingly, the cold start runs were much more successful in
finding a single free energy minimum at $T / J$ between 1.3125 and 1.40625.
In the majority of samples, all of the cold start initial conditions were
converging to the same free energy minimum.  This was no longer true for
$T / J < 1.3125$, and thus results for these lower temperatures are not
reported.

For each sample at each value of $T / J$, the data shown in the figure were
obtained by averaging $S ( k )$ over 16 spin states taken at intervals of
51,200 Monte Carlo steps per spin (MCS), after letting the sample equilibrate.
Thus the length of each data-gathering run was 768,000 MCS.  The values shown
for $S ( k )$ are obtained by averaging over all of the values of $\vec{\bf k}$
which have length $k$.  This multiplicity, which is not the same for all values
of $k$, was not taken into account in calculating the error bars shown in the
figures.  Thus each error bar shown in the figures represents approximately
three standard deviations.  The procedure used for the Type B samples was
essentially identical.

\begin{figure}
\includegraphics[width=3.4in]{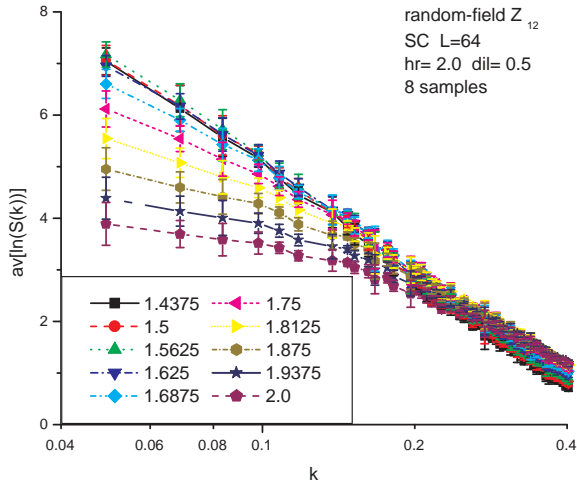}
\caption{\label{Fig.2}(color online) Average over 8 samples of $\ln(S(k))$
for $64 \times 64 \times 64$ lattices with $h_r = 2$ and $p = 0.5$ at
various temperatures. The error bars indicate approximately three standard
deviation statistical errors, and the $x$-axis is scaled logarithmically.}
\end{figure}

In Fig.~2 we show the corresponding data for the Type B samples, for $T / J$
between 1.4375 and 2.00.  The data for $T / J > 1.5625$ are taken from the
hot start runs, and the data for lower values of $T / J$ are taken from the
cold start run which gave the lower average value for the energy,
$\langle E \rangle$.  It was also true for these Type B samples that the hot
start samples were unable to equilibrate on accessible time scales for
$T / J \le 1.5625$.

\begin{figure}
\includegraphics[width=3.4in]{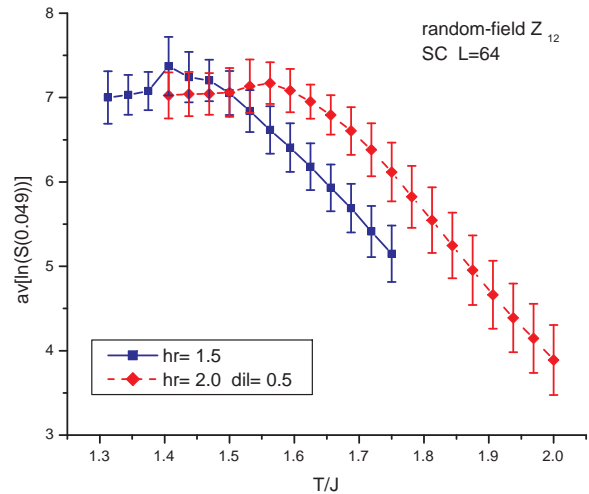}
\caption{\label{Fig.3}(color online) Average over 8 samples of $\ln(S(\pi
/ 64))$ for $64 \times 64 \times 64$ lattices as a function of $T / J$.
The error bars indicate approximately three standard deviation statistical
errors.}
\end{figure}

We can define $T_c$ in a number of ways.  For instance, it can be defined
as the highest temperature for which we have been unable to equilibrate the
hot start runs.  It is also the temperature for which the equilibration
time for the cold start runs becomes very long.  The fact that
equilibration times get very long at the same $T_c$ for both the hot start
and the cold start initial conditions that the behavior of this model can
not be explained by the usual hysteresis which is associated with a normal
first-order phase transition.  As was discussed in somewhat more detail in
the earlier work,\cite{Fis07} the author believes that this phase transition
should be described by a theory of the Anderson-Yuval\cite{AY71} type, where
the order parameter jumps to zero discontinuously at $T_c$.  A phase
transition of this type is also believed to occur in the {\it k}-core
percolation model,\cite{SLC06} which has been suggested as model for the
ordinary glass transition.

The qualitative behavior which we see in Fig.~1 and Fig.~2 is that the
average over samples of $\ln(S(k))$ shows a peak at small $k$ which grows
in size as we lower $T$, until it reaches a maximum at $T_c$.  The value
of $T_c / J$ is about 1.40625 for the Type A samples and 1.5625 for the
Type B samples.  Upon lowering $T$ below $T_c$, the size of the peak becomes
somewhat smaller.  This is seen more clearly in Fig.~3, which shows the
temperature dependence of av[$\ln(S(\pi / 64))$] for both types of samples.
(The value of $\pi / 64$ is approximately 0.049.)

For $T > T_c$ both types display a substantial range of $T$ over which
$S(\pi / 64)$ is increasing exponentially as $T$ is lowered, and this
increase slows very close to $T_c$.  This kind of temperature dependence is
well known phenomenologically\cite{WLF55} to be a characteristic of glassy
freezing behavior.  However, the reader should keep in mind that one expects
the usual glass transition to correspond to $n = 3$ spins, and not the
$n = 2$ spins studied here.  If one generalizes Eqn.~(1) to the $n = 3$ case,
substantially more computing resources would be required to make the
equivalent calculation feasible.

Below $T_c$, $S(\pi / 64)$ first decreases, and then becomes approximately
constant, at least over the range of $T$ shown in the figure, where the
samples are believed to be equilibrated.  This behavior below $T_c$ is
consistent with the existence of a $\delta$-function peak in $S$ at $k = 0$
in the limit $L \to \infty$, indicative of true long-range order.  One
would not expect to see this behavior if there were only power-law
correlations in $S ( k )$, with no $\delta$-function peak.

The reader should understand that we are not claiming the existence of some
mathematical error in the analysis\cite{Fish97} of the elastic glass, which
leads to the power-law correlated Bragg glass phase.  That analysis begins
with the Larkin Hamiltonian,\cite{Lar70} which approximates the spin-exchange
term of Eqn.~(1) with a harmonic interaction.  What we are saying is that the
long-range behavior of the Larkin Hamiltonian is different from that of
Eqn.~(1), at least for the types of $P ( h_i )$ distributions we have studied.

If we fit the data for av[$\ln(S( k ))$] at $T_c$ over the range
$\pi / 64 \le k \le \pi / 8$, as shown in Fig.~1 and Fig.~2, we find good
straight line fits, with slopes of -3.20(3) for the Type A samples and
-3.10(2) for the Type B samples.  Because both of these slopes are more
negative than -3, it is clearly not correct to extrapolate these fits to
$k = 0$.  $S(k)$ satisfies a sum rule, and thus nonintegrable singularities
cannot occur.  Thus if we were able to obtain data using these $P ( h_i )$
distributions for even larger values of $L$, we would expect to find that
the behavior of av[$\ln(S( k ))$] at $T_c$ for small enough nonzero $k$
would look like the data for the case $h_r = 2$ with $p = 0$ found in the
earlier work.\cite{Fis07}

In this work we have performed Monte Carlo studies of the 3D RFXYM
on $L = 64$ simple cubic lattices, with isotropic random field
distributions of two types: $h_r$ = 1.5 with $p$ = 0, and $h_r$ = 2 with
$p$ = 0.5.  We present results for the structure factor, $S ( k )$, at a
sequence of temperatures.  In agreement with prior results\cite{Fis07}
on weaker and stronger random fields, we argue that our results indicate
a phase transition into a long-range ordered low temperature phase, with
a jump in the order parameter at $T_c$.  The behavior above $T_c$ strongly
resembles well known phenomenology for glassy phase transitions.\cite{WLF55}

\begin{acknowledgments}
The author thanks the Physics Department of Princeton University for
providing use of computers on which the data were obtained.

\end{acknowledgments}


\begin{thebibliography}{49}

\bibitem{Lar70}
A. I. Larkin, Zh. Eksp. Teor. Fiz. {\bf 58}, 1466 (1970) [Sov. Phys.
JETP {\bf 31}, 784 (1970)].
\bibitem{GH96}
M. J. P. Gingras and D. A. Huse, Phys. Rev. B {\bf 53}, 15193
(1996).
\bibitem{Fis97}
R. Fisch, Phys. Rev. B {\bf 55}, 8211 (1997).
\bibitem{SJSTAT98}
P. J. Steinhart {\it et al.}, Nature {\bf 396}, 55 (1998).
\bibitem{IM75}
Y. Imry and S.-K. Ma, Phys. Rev. Lett. {\bf 35}, 1399 (1975).
\bibitem{Grin76}
G. Grinstein, Phys. Rev. Lett. {\bf 37}, 944 (1976).
\bibitem{AIM76}
A. Aharony, Y. Imry and S.-K. Ma, Phys. Rev. Lett. {\bf 36}, 1364
(1976).
\bibitem{PS82}
G. Parisi and N. Sourlas, Nucl. Phys. {\bf B206}, 321 (1982).
\bibitem{Imb84}
J. Z. Imbrie, Phys. Rev. Lett. {\bf 53}, 1747 (1984).
\bibitem{BK87}
J. Bricmont and A. Kupiainen, Phys. Rev. Lett. {\bf 59}, 1829
(1987).
\bibitem{Grif66}
R. B. Griffiths, Phys. Rev. {\bf 152}, 240 (1966).
\bibitem{Fis07}
R. Fisch, Phys. Rev. B {\bf 76}, 214435 (2007).
\bibitem{AY71}
P. W. Anderson and G. Yuval, J. Phys. C {\bf 4}, 607 (1971).
\bibitem{SLC06}
J. M. Schwarz, A. J. Liu and L. Q. Chayes, Europhys. Lett. {\bf 73},
560 (2006).
\bibitem{WLF55}
M. L. Williams, R. F. Landel and J. D. Ferry, J. Am. Chem. Soc.
{\bf 77}, 3701 (1955).
\bibitem{Fish97}
D. S. Fisher, Phys. Rev. Lett. {\bf 78}, 1964 (1997).


\end{thebibliography}


\end{document}